\documentclass[sigconf,authorversion=true]{acmart}

\usepackage{booktabs} 
\usepackage{balance}
\usepackage{algorithm,algpseudocode}
\usepackage{amsmath}

\usepackage[caption=false,font=footnotesize]{subfig}
\usepackage{fancyhdr}
\usepackage{mdwlist}
\usepackage{multirow}
\usepackage{amsmath}        

\graphicspath{{incl/}}

\makeatletter
\def\blfootnote{\xdef\@thefnmark{}\@footnotetext}
\makeatother

\usepackage{setspace}
\def\smallerspacecaption{\vspace{-2mm}}

\definecolor{gray}{gray}{0.9}

\newcommand{\drop}[1]{}

\begin{document}

\title[Best of Both Worlds: Integration of Split Man.\ and Camouflaging for 3D ICs]{Best of Both Worlds: Integration of Split Manufacturing and Camouflaging into a Security-Driven CAD Flow for 3D ICs}

\author{Satwik Patnaik, Mohammed Ashraf, Ozgur Sinanoglu, and Johann Knechtel}
 \affiliation{%
   \institution{Tandon School of Engineering, New York University, New York, USA}
   \institution{Division of Engineering, New York University Abu Dhabi, United Arab Emirates}
 }
 \email{{sp4012, ma199, ozgursin, johann} @nyu.edu}

\renewcommand{\shortauthors}{S.\ Patnaik et al.}

\begin{abstract}

With the globalization of manufacturing and supply chains, ensuring the security and trustworthiness of ICs has become an urgent challenge.
Split manufacturing (SM) and layout camouflaging (LC) are promising techniques to protect the intellectual property (IP)
of ICs from malicious entities during and after manufacturing (i.e., from untrusted foundries and reverse-engineering by end-users).
In this paper, we strive for ``the best of both worlds,'' that is of SM and LC.
To do so, we extend both techniques towards 3D integration, an up-and-coming design and manufacturing paradigm based on stacking and
interconnecting of multiple chips/dies/tiers.

Initially, we review prior art
and their limitations.
We also put forward a novel, practical threat model of IP piracy which is in line with the business models of present-day
design houses.
Next, we discuss how 3D integration is a naturally strong match to
	combine SM and LC.
We propose a security-driven CAD and manufacturing flow for face-to-face (F2F) 3D ICs, along with obfuscation of interconnects.
Based on this CAD flow, we conduct comprehensive experiments on DRC-clean layouts. Strengthened by an extensive security analysis (also
		based on a novel attack to recover obfuscated F2F interconnects), we argue that entering the next, third dimension is
eminent for effective and efficient IP protection.

\end{abstract}

\copyrightyear{2018} 
\acmYear{2018} 
\setcopyright{acmcopyright}
\acmConference[ICCAD '18]{IEEE/ACM INTERNATIONAL CONFERENCE ON COMPUTER-AIDED DESIGN}{November 5--8, 2018}{San Diego, CA, USA}
\acmBooktitle{IEEE/ACM INTERNATIONAL CONFERENCE ON COMPUTER-AIDED DESIGN (ICCAD '18), November 5--8, 2018, San Diego, CA, USA}
\acmPrice{15.00}
\acmDOI{10.1145/3240765.3240784}
\acmISBN{978-1-4503-5950-4/18/11}

\maketitle

\renewcommand{\arraystretch}{.94}

\section{Introduction}
\label{sec:introduction}

On the one hand, design practices by the industry attach importance to optimize for power, performance, and area
(PPA) at the level of physical design or
design architecture (e.g., cache hierarchies, speculative execution).
On the other hand, researchers have demonstrated powerful attacks (e.g., \emph{Spectre}~\cite{Kocher2018spectre}
		or side-channel leakage~\cite{lerman18})
which leverage these very practices and optimization steps.
Apart from such concerns regarding the security and trustworthiness of hardware at runtime, protecting the hardware itself from
threats such as intellectual property (IP) piracy, illegal overproduction or insertion of hardware Trojans is another
challenge~\cite{BRS17}.
Various design and manufacturing schemes have been put forth over the last
decade, e.g., ranging from logic locking~\cite{yasin17_CCS,shamsi18_TIFS}, layout
camouflaging~\cite{rajendran13_camouflage,wang16_MUX,collantes16,nirmala16,
	Akkaya2018,chen15,patnaik17_Camo_BEOL_ICCAD,patnaik18_GSHE_DATE}, to split
manufacturing~\cite{
	rajendran13_split,wang17,patnaik_ASPDAC18,patnaik18_SM_DAC,wang18_SM,magana17,sengupta17_SM_ICCAD}.
The common theme among these techniques is that
they incorporate security as a critical design parameter besides the traditional PPA metrics.

Independent of hardware security, 3D integration has made significant progress over recent years. 3D integration is to
stack and interconnect multiple chips/dies/tiers, thereby promising
to overcome the scalability bottleneck (``More-Moore''), which is further exacerbated
by challenges for pitch scaling, routing congestion, process variations, et cetera~\cite{knechtel16_Challenges_ISPD,EF16}.
Recent studies and prototypes show that 3D integration can indeed offer significant benefits over conventional 2D
chips~\cite{peng17, jung14, kim12_3dmaps}.
Besides, 3D integration advances manufacturing capabilities by various means such as parallel handling of wafers, higher yields due to smaller
outlines of individual chips, and heterogeneous integration (``More-than-Moore'')~\cite{radojcic17}.

\begin{figure}[tb]
\centering
\includegraphics[width=.88\columnwidth]{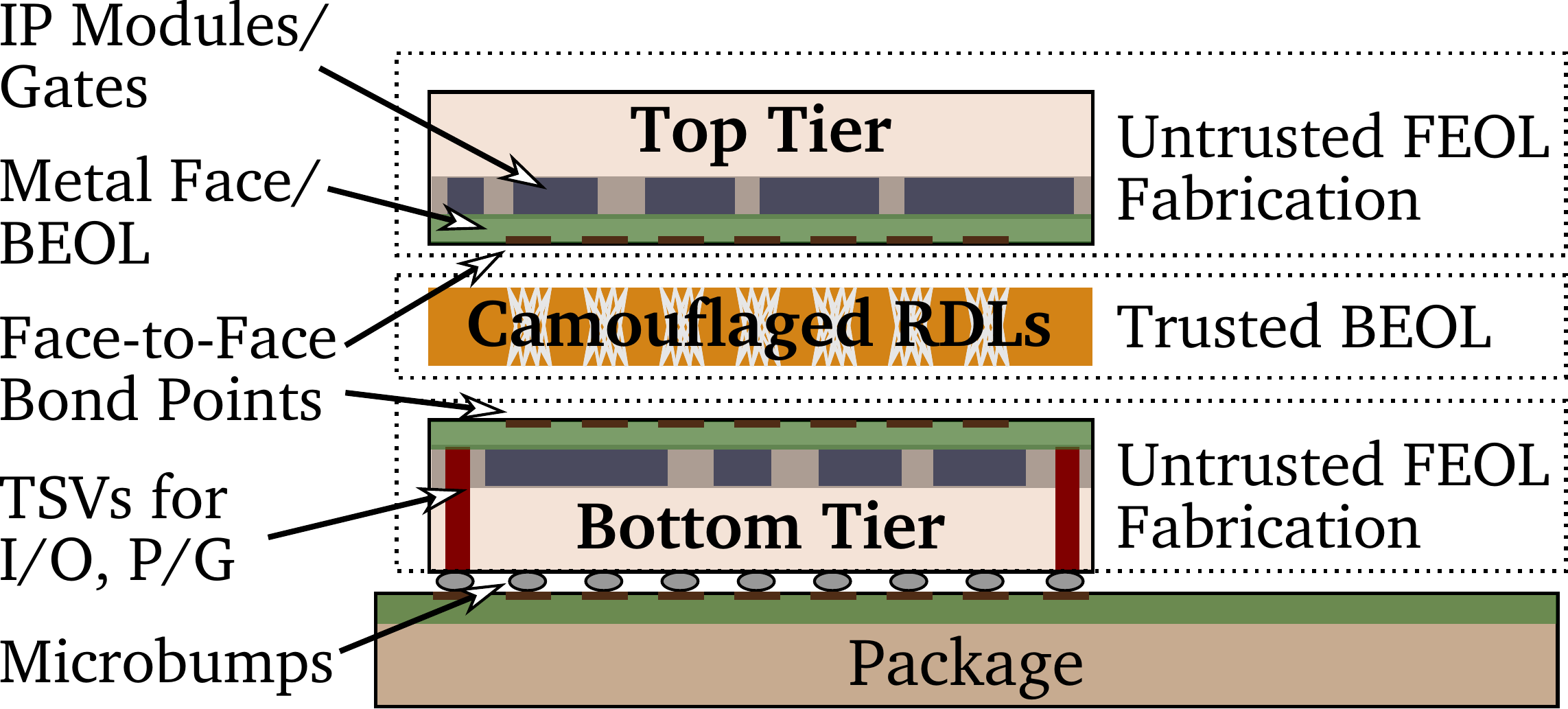}
\smallerspacecaption
\caption{Our security-driven scheme for 3D integration, focused on face-to-face (F2F) 3D ICs.
	Through-silicon vias (TSVs) are for external connections, and redistribution layers (RDLs)
	for internal connections.
	We advance split manufacturing for
	untrusted FEOL fabrication
		along with trusted camouflaging of
		RDLs---both techniques are a natural match for taking IP protection to the next, third dimension.
\label{fig:scheme}
}
\smallerspacecaption
\end{figure}

In this paper, we propose and evaluate a security-driven CAD flow for 3D ICs. We argue that 3D integration is an excellent candidate for IP
protection, and we demonstrate that by combining layout camouflaging (LC) and split manufacturing (SM)
naturally into one scheme
(Fig.~\ref{fig:scheme}).
This paper can be summarized as follows:

\begin{itemize}

\item Initially, we review state-of-the-art approaches for LC and SM.
We compare and contrast these schemes with regards to their security guarantees, shortcomings, and impact on PPA.

\item Next, we put forward a practical threat model which is in line with the present-day business models of design houses.
This model necessitates both LC and SM in conjunction.

\item Most importantly, we demonstrate how 3D integration can help to achieve the ``best of both worlds,'' by combining the features of LC
and SM.
	Thus, our scheme allows
	to inherently protect against IP piracy conducted by malicious entities during fabrication (untrusted foundries) and after
	fabrication (untrusted end-users).
The key idea is to ``3D split'' the design into two tiers and to obfuscate the interconnects between those tiers.
Towards this end,
we propose a security-driven CAD and manufacturing flow for face-to-face (F2F) 3D ICs.

\item We implement our CAD flow using \emph{Cadence Innovus}.  We conduct a thorough analysis of DRC-clean layouts tailored for F2F 3D
integration, and we contrast with the prior art of LC or SM (targeting on 2D/3D ICs) wherever applicable.

\item We present an extensive security analysis, underpinned by a novel proximity-centric attack on our security-driven 3D integration
scheme. We provide both analytical and empirical data to showcase the resilience of our proposed schemes.

\end{itemize}

\section{Background}
\label{sec:background}

\subsection{Layout Camouflaging}
\label{sec:layout_camo}

Camouflaging (LC) is a layout-level technique to foil an adversary's efforts for correctly inferring the design functionality while
reverse engineering some chip. LC is accomplished during manufacturing by ($i$) dissolving optically distinguishable traits of standard
cells, e.g., using look-alike gates~\cite{rajendran13_camouflage} or secretly configured MUXes~\cite{wang16_MUX}, ($ii$) using selective
doping implantation for threshold-voltage-based obfuscation~\cite{collantes16,nirmala16,
		Akkaya2018}, or ($iii$) rendering the
BEOL wires and/or vias resilient against reverse engineering~\cite{chen15,patnaik17_Camo_BEOL_ICCAD}.

\begin{figure}[b]
	\smallerspacecaption
	\captionsetup[subfigure]{labelformat=empty}
\centering
	\subfloat[]{
\includegraphics[width=.48\columnwidth]{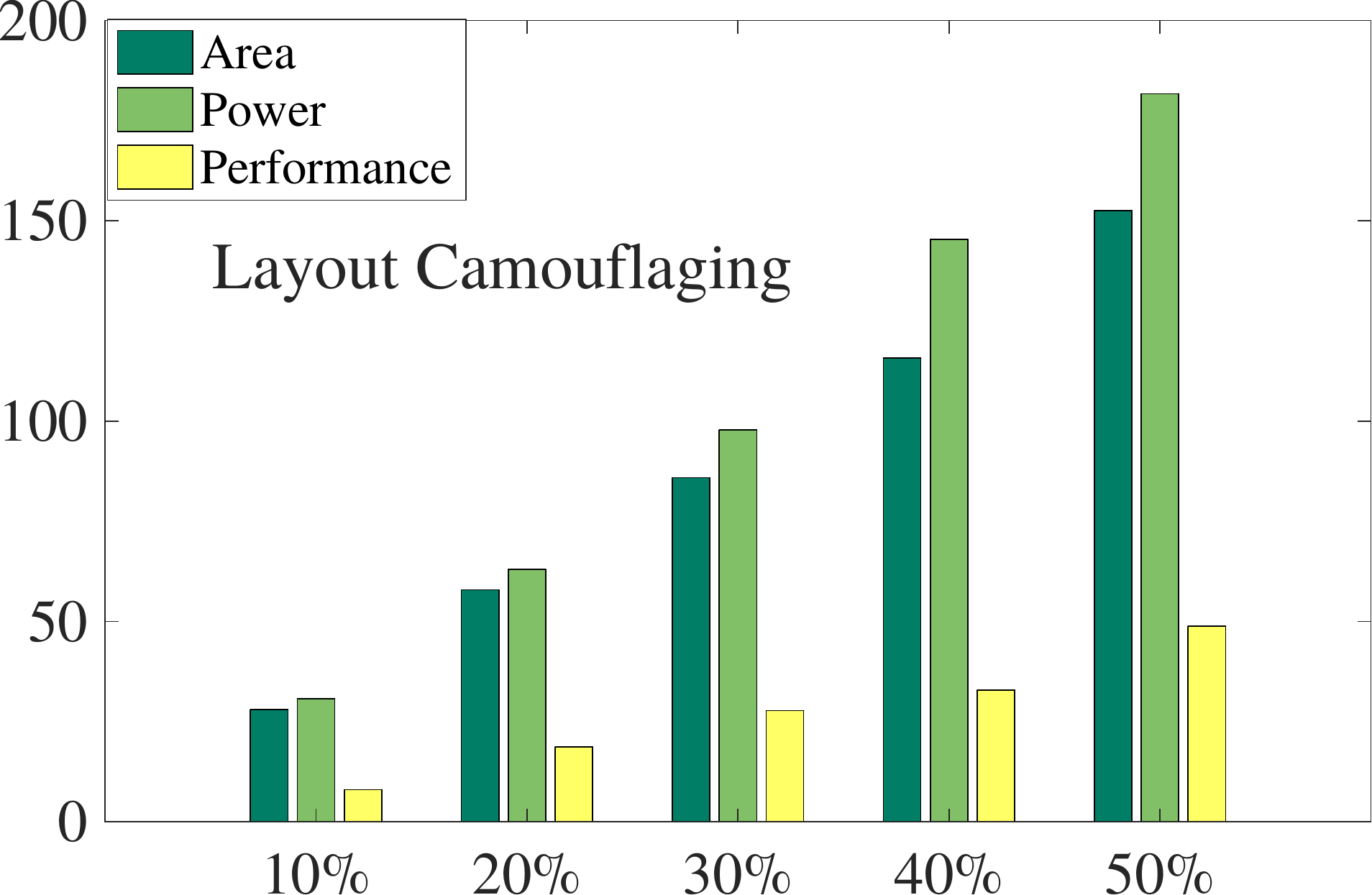}
	}
\hfill
	\subfloat[]{
\includegraphics[width=.48\columnwidth]{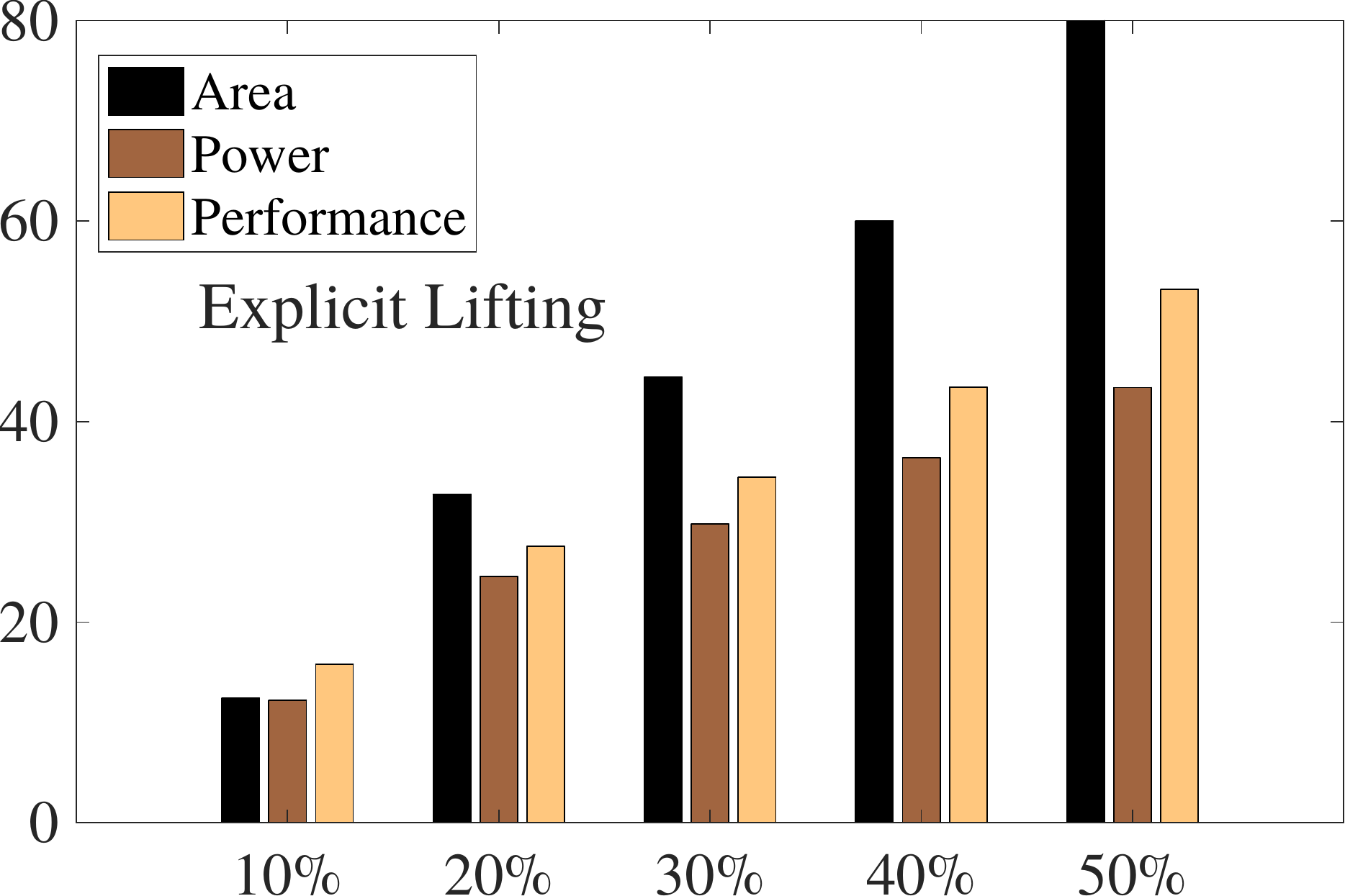}
	}
	\smallerspacecaption
\smallerspacecaption
	\smallerspacecaption
\caption{PPA cost in \% for look-alike camouflaging~\cite{rajendran13_camouflage} (left) and explicit lifting of randomly selected wires to
	M8 (right). Results are averaged across
	\emph{ITC-99} benchmarks. For LC (left),
	the impact on power and area is substantial, given that the NAND-NOR-XOR structure in~\cite{rajendran13_split} incurs 4$\times$ and
		5.5$\times$ more area and power compared to a regular 2-input NAND gate.
		For SM (right), the cost
	for area is severe; that is because routing resources are relatively scarce for M8 (pitch = 0.84$\mu m$), and 
		lifting of wires occupies further resources, which can only be obtained by enlarging the chip outlines.
\label{fig:naive_lifting_overheads}
\label{fig:LC_overheads}
}
\end{figure}

Existing schemes can incur significant PPA overheads once LC is applied for large parts of the design.
For example for~\cite{rajendran13_camouflage},
camouflaging 50\% of the design results in $\approx$150\% overheads for power and area, respectively
(Fig.~\ref{fig:LC_overheads}).
Other emerging schemes such as 
threshold-voltage-based LC can suffer from massive PPA overheads as well; see Sec.~\ref{sec:results} for more comparative results.
Also note that most schemes require alterations to the FEOL manufacturing process, which can be costly.
There, since camouflaging builds the secret for IP protection, the commissioned FEOL fab has to be \emph{trusted}.

\subsection{Split Manufacturing}
\label{sec:split_manufacturing}

Split manufacturing (SM) offers an interesting solution to safeguard the design IP during manufacturing time.
   Most commonly SM means that
the device layer and few lower metal layers (front-end-of-line, FEOL) are fabricated using a high-end, 
   potentially \emph{untrusted} foundry, whereas the remaining interconnects (back-end-of-line, BEOL) are grown
   on top of the FEOL wafer by a \emph{trusted} facility.
   The security promise lies in the fact that the
   \emph{untrusted} foundry only holds a part of the overall design, making it difficult to infer the complete design functionality, and thereby
   hindering an adversary from IP piracy or targeted insertion of hardware Trojans.

Existing CAD tools, however, due to their focus on design closure (and their so-far agnostic view on security), tend to leave 
hints for an FEOL-based adversary. To honor PPA, for example, to-be-connected cells are typically placed close to each other.
	Hence,
      Rajendran~\emph{et al.}~\cite{rajendran13_split} proposed a so-called proximity attack which models
this principle to infer the missing BEOL connections.

Various placement-centric~\cite{wang18_SM,sengupta17_SM_ICCAD,patnaik18_SM_DAC} and/or
routing-centric~\cite{wang17,patnaik_ASPDAC18,patnaik18_SM_DAC} schemes have been proposed recently, which all aim to counter the efforts of
various iterations of proximity attacks~\cite{wang18_SM,magana17}.
Among those defense schemes, lifting of wires above the split layer
is an intuitive way to obfuscate the IP.
That is, the revealing or critical wires
(as selected by the designer)
are lifted, e.g., with the help of
routing pins in higher layers.
In our exploratory
experiments on randomized lifting of nets (Fig.~\ref{fig:naive_lifting_overheads}), we observe steady increases in PPA cost.  As with LC,
	    more comparative results are given in Sec.~\ref{sec:results}.

\subsection{3D Integration and CAD Flows}
\label{sec:limitations_3D_CAD}

3D integration can be classified into four flavors:
(1) through-silicon via (TSV)-based 3D ICs, where chips are fabricated separately and then stacked, with inter-chip connections being
realized by TSVs connected to metal layers;
(2) face-to-face (F2F) stacking, where two tiers are fabricated separately and then bonded together at their metal faces;
(3) monolithic 3D ICs, where multiple tiers
are manufactured sequentially, with inter-tier connects based on regular metal vias;
(4) 2.5D integration, where chips are fabricated separately and then bonded to a system-level interconnect
carrier, the interposer. Each option has its scope, benefits and drawbacks, and requirements for CAD and manufacturing
processes~\cite{knechtel16_Challenges_ISPD,knechtel17_TSLDM}.
	
F2F stacking has arguably emerged as most promising (along with monolithic 3D ICs);
various studies are actively streamlining efforts for commercial adoption~\cite{chang16,ku18,peng17,peng15_F2F}.
The principal goal of these studies is to optimize for PPA and the microarchitecture, not hardware security.
More specifically, prior CAD flows carefully trade off intra-tier wiring with vertical interconnects across tiers.
While the latter is the key feature of 3D integration, an overly large number of crossings/cuts has a significant impact on PPA as well.
As we will explain in Sec.~\ref{sec:security} in more detail, however, a large number of cuts is mandatory for a strong resilience against IP
piracy.

\section{A Practical Threat Model}
\label{sec:threat_model}

\begin{figure}[tb]
\centering
\smallerspacecaption
\includegraphics[width=.85\columnwidth]{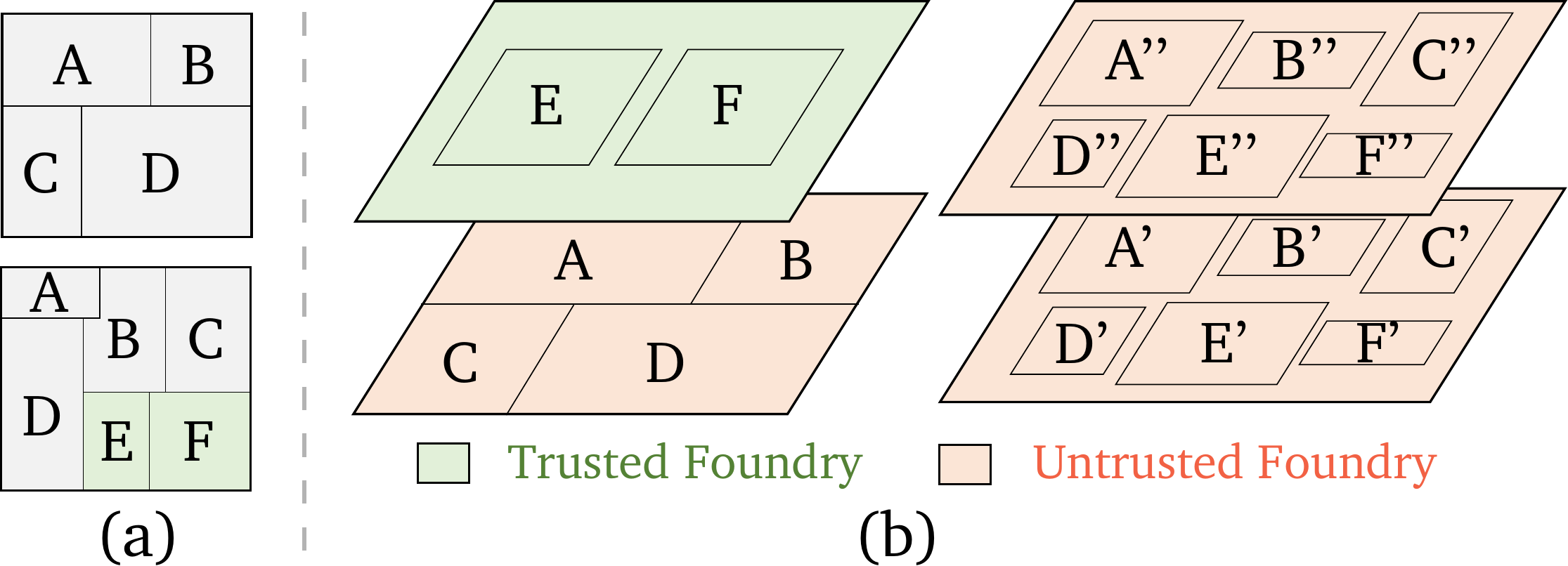}
\smallerspacecaption
	\caption{(a) Current chip version (top) versus new chip version (bottom).
		In the new version, the IP modules \emph{E} and \emph{F} are entirely new, while the
		other modules are revised and/or reshaped.
			(b) Foundry scenarios for our security-driven 3D integration scheme.
			For both tiers manufactured by an untrusted foundry (right), IP modules can be split up.
    \label{fig:foundry_options}
	}
\smallerspacecaption
\smallerspacecaption
\end{figure}

Here we put forward a novel, practical threat model for IP piracy, which is in line with the business models of present-day design houses.
	Consider the following
	scenario.
A design house commissions an untrustworthy foundry to manufacture their newest version of some chip.
This new version is typically extended
from previous versions of the chip (Fig.~\ref{fig:foundry_options})---the reuse of IP modules and the re-purposing of proven
architectures are well-known principles.
Hence, the previous
versions of the chip can be obtained from the market.
For example, think of the flagship iPhone\textsuperscript{\textregistered} by Apple\textsuperscript{\textregistered}. The iPhone
7,
	based on the A10 chip,
	was launched in September 2016, and the
iPhone X, based on the successor chip A11,
was launched in September 2017---both chips are available in the market.
In this scenario, it is intuitive that recovering the new IP can become significantly less challenging for the potentially untrustworthy fab.
In case the same fab was already commissioned for the previous version, it readily holds that prior layout; otherwise it can reverse
engineer the layout from chips bought in the market. In any case, the adversaries can compare that prior layout with the
new layout, to locate and focus on those parts which are different and unique.

Now, the conclusion for this thought experiment is that \emph{both SM and LC are required for manufacturing of all different chip versions}.
LC is required to prevent reverse engineering of the current layout by any other fab commissioned for later chip
versions; SM is necessary to prevent the fab which is manufacturing the current version (and
which is also tasked to implement LC) from readily inferring the complete layout of the current version.
Prior art can only account for this practical threat model by applying both SM \emph{and} LC, which can exacerbate
the overheads and shortcomings as discussed in Sec.~\ref{sec:background}.
Next, we outline our scheme to combine SM and LC naturally while leveraging 3D integration.

\section{3D Integration: Our Concept for IP Protection in the Next Dimension}
\label{sec:concept}

The primary advancement we propose for SM is to ``3D split'' the design into multiple tiers. That is, unlike regular SM in 2D where the
whole layout is split into FEOL and BEOL, here we split the layout itself into two parts.
These two parts are manufactured as separate
chips and then stacked and interconnected, in this paper based on the F2F flow.
Our work is the \emph{first} which demonstrates such a natural extension of SM.\footnote{We acknowledge that the idea for 3D SM was
envisioned in 2008 by Tezzaron~\cite{tezzaron08}.
Also, there are studies hinting at the benefits of 3D integration for SM~\cite{dofe17,gu16,imeson13,xie17,valamehr13}, but all have
shortcomings or cover different scenarios: Dofe \emph{et al.}~\cite{dofe17} and Gu \emph{et al.}~\cite{gu16} remain on the
conceptional level; Xie \emph{et al.}~\cite{xie17} and Imeson \emph{et al.}~\cite{imeson13}
consider 2.5D integration where only wires are hidden from the untrusted foundry;
	Valamehr \emph{et al.}~\cite{valamehr13} propose to stack customized monitoring circuitry on top
	of untrustworthy chips, i.e., they leverage 3D integration for runtime monitoring, not for IP protection.}
We suggest that 3D SM can be done either by different foundries or by one foundry
(Fig.~\ref{fig:foundry_options}):

\begin{enumerate}
\item \emph{Different trusted and untrusted foundries:}
Here we delegate the manufacturing to one low-end but trusted and one high-end but
untrusted foundry, both with FEOL/BEOL capabilities. A chip
company may have significantly more options to commission a trustworthy foundry (or even manufacture in-house) in case the sought-after
technology node is old yet still widely available, e.g., 180nm.
While keeping one design part exclusively with a trusted foundry is promising security-wise, the practicality of this option seems limited.
\item \emph{Untrusted foundries/foundry:} Here we commission only high-end but untrusted foundries for both parts/chips.
This way, we may benefit from the latest technology node but, naturally, have to split the design in such a way
that the foundries cannot readily infer the whole layout, even when they are colluding.
Once such strong protection is in place,
it is economically more reasonable to commission only one foundry.
\end{enumerate}

\subsection{Different Trusted and Untrusted Foundries}
\label{sec:diff_found}

The commissioning of several foundries with different technologies and trust levels
has some critical implications as follows.

First, regarding the practical threat model, it is straightforward to assign the new IP exclusively to the chip manufactured by the
trusted foundry.
	As for the resilience of this inherently secure 3D SM scheme, there is no generic attack model in the literature yet which can
	account for this scenario, that is, when given only one part of the layout how to infer the missing connections \emph{and} gates. We
	believe that a corresponding ``black-box'' attack would be very challenging, but we suggest that the community may consider it.
   
Second, due to the different pitches for different technologies,
only a fraction of the design can be delegated to the low-end chip. That is at least as long as
(a) the high-end chip shall have reasonable utilization for cost efficiency and (b) the outlines of both
chips
shall remain the same, which is a common requirement for 3D stacking.

Third, the overall power and performance is dominated by the low-end chip, where other factors such as parasitics may further exacerbate the
overheads in practice~\cite{peng17}.

\begin{figure}[tb]
\centering
\includegraphics[width=.8\columnwidth]{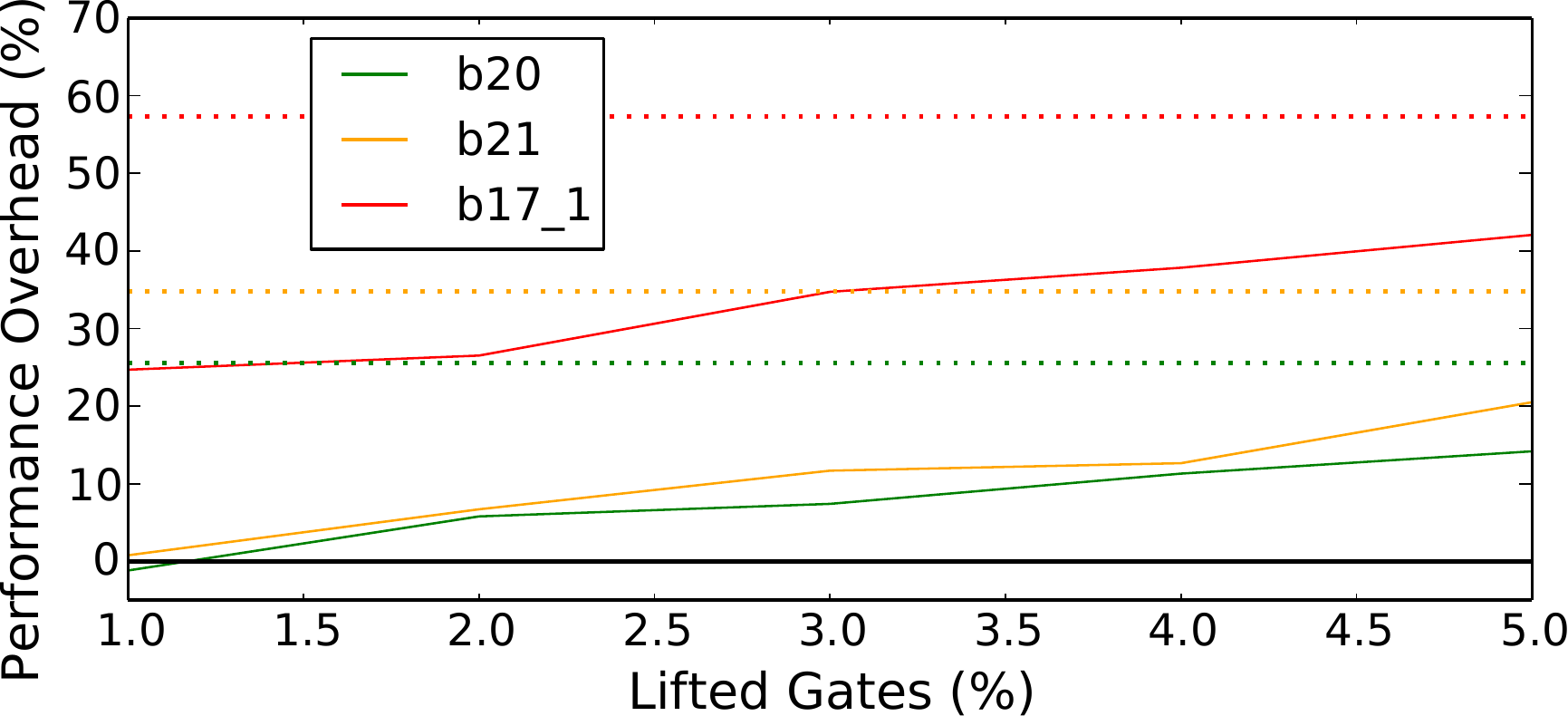}
\smallerspacecaption
\caption{Performance degradation when lifting gates in 3D from the 45nm tier to the 180nm tier.
		The dotted lines indicate the critical-path delays for the 180nm 2D baseline setup.
			See also Table~\ref{tab:basic_PPA_two_tech} and Footnote~\ref{fn:two_tech}.
\label{fig:two_tech_delay}
}
\smallerspacecaption
\smallerspacecaption
\end{figure}

In our exploratory experiments, we gauge the capabilities for such 3D SM,
assuming a trusted 180nm foundry and an untrusted 45nm foundry.
More specifically, we leverage
the \emph{OSU} libraries~\cite{OSU_PDK}.
Their libraries hold the same number, type, and strengths of cells; this guarantees a fair comparison since CAD tools cannot leverage
different versions of cells.
\emph{Synopsys DC} was used for synthesis and place and
route was performed using \emph{Cadence Innovus 17.1}.
PPA results for an aggressive timing closure of the 2D baseline setup are given in
Table~\ref{tab:basic_PPA_two_tech}.\footnote{\label{fn:two_tech}
The node 45nm is four generations away from 180nm, and delays improve by $\approx$30\% per generation~\cite{borkar1999design};
surprisingly, delay degradations for the \emph{OSU} 180nm library are notably below such expectations.
We believe that this is due the academic nature of
the library. At the point of writing, we had no access to different commercial libraries.
In any case, the key findings for our experiments remain valid.
That is because once the delay numbers in the library would be revised,
and still assuming the same types of cells are used, the overall delay would merely scale up linearly.}
For the heterogeneous F2F 3D setup (Fig.~\ref{fig:two_tech_delay}),
we observe some performance degradation as we lift more gates to the low-end tier.
Also, note from Table~\ref{tab:basic_PPA_two_tech} that
area (and power) cost is $\approx$12X (and 9X) when contrasting 180nm to 45nm.
To maintain a balanced utilization for both tiers, these correlations imply that one should not lift more than $\approx$8\% of the gates to
the low-end tier.
While such small-scale lifting provides a reasonable performance gain, especially from the
perspective of commissioning only the 180nm foundry, it may not be enough to cover all the sensitive IP.

In short, we find
that leveraging different foundries has practical limitations.
Delegating more than $\approx$8\% of the gates
to the low-end foundry
is ineffective, especially when considering that 
this foundry has to implement LC as well, which incurs further cost.

\begin{table}[tb]
\centering
\scriptsize
\setlength{\tabcolsep}{0.33em}
\caption{Timing-aggressive 2D baselines, based on the \emph{OSU} libraries~\cite{OSU_PDK}.
	All layouts are DRC clean. Area is in $\mu m^2$, power in $mW$, and delay in $ns$.
			See also Footnote~\ref{fn:two_tech}.
}
\smallerspacecaption
\smallerspacecaption
\begin{tabular}{|c|c|c|c|c|c|c|c|c|}
\hline
\multirow{2}{*}{\textbf{Benchmark}} & \multicolumn{4}{|c|}{\textbf{45nm}} & \multicolumn{4}{|c|}{\textbf{180nm}} \\
\cline{2-9}
 &
 \textbf{\# Instances} &
 \textbf{Area} &
 \textbf{Power} &
 \textbf{Delay} &
 \textbf{\# Instances} &
 \textbf{Area} &
 \textbf{Power} &
 \textbf{Delay} \\
 \hline \hline
\emph{b17\_1} &  
14,850 & 32,770.28  & 8.85  & 2.29  & 14,711  & 417,416 & 71.54 & 3.59 \\ \hline
\emph{b20} &  
6,959 & 15,549.31  & 8.12  & 2.87  & 7,521  & 216,168 & 97.94 & 3.6 \\ \hline
\emph{b21} &  
7,327 & 16,096.05  & 8.79  & 2.88 & 7,060  & 203,216 & 85.66 & 3.89 \\ \hline
\end{tabular}
\label{tab:basic_PPA_two_tech}
\smallerspacecaption
\end{table}

\subsection{Untrusted Foundries}
\label{sec:untrusted_foundry}

Engaging with several untrusted foundries offering the same technology (or one untrusted foundry) also holds some key implications.

First, power and performance of such ``conventional 3D ICs'' can be expected to excel those of the different-technology scenario above.
In fact, folding (or splitting) of 2D IP modules within 3D ICs has been successfully demonstrated for some time~\cite{lin16,jung14,jung17,peng15_F2F}, albeit
without IP piracy in mind.
Hence, savings from the folding of IP modules may provide some margin for a defense scheme, but we show in the remainder of this work that
this margin naturally depends on the design and the measures applied for protection.

Second, although
	IP modules can be folded/split across tiers, which may mislead a reverse-engineering attacker, both
tiers are still manufactured by untrusted foundries.
This fact implies that LC schemes targeting on the device level
\emph{cannot} help to protect the IP from adversaries in those foundries.
Interestingly, there is another LC flavor emerging,
	that is the obfuscation of
interconnects~\cite{cocchi14,chen15,vijayakumar16,patnaik17_Camo_BEOL_ICCAD}.
Chen \emph{et al.}~\cite{chen15} consider real and dummy vias using magnesium, Mg and magnesium-oxide, MgO, respectively.
They demonstrate that real Mg vias oxidize quickly into MgO and, hence, can become indistinguishable from the other MgO dummy vias during
reverse engineering.
Hwang \emph{et al.}~\cite{hwang12_transient_electronics} have shown that Mg and MgO dissolve quickly when
surrounded by fluids, which is inevitable in etching procedures applied for reverse engineering.
Thus, without loss of generality, we assume our interconnect obfuscation to be based on the use of Mg/MgO vias.

We argue that the obfuscation of interconnects is a natural match for F2F 3D ICs---in between the tiers, further
redistribution layers (RDLs) can be purposefully manufactured for obfuscation.
Doing so only requires a trustworthy BEOL facility, which is a practical assumption given that BEOL fabrication is
much less demanding than FEOL fabrication, especially for higher metal layers (RDLs reside between the F2F bonds which themselves are at higher
		layers).

\section{Methodology}
\label{sec:methodology}

\begin{figure}[tb]
\smallerspacecaption
\smallerspacecaption
\centering
\includegraphics[width=.95\columnwidth]{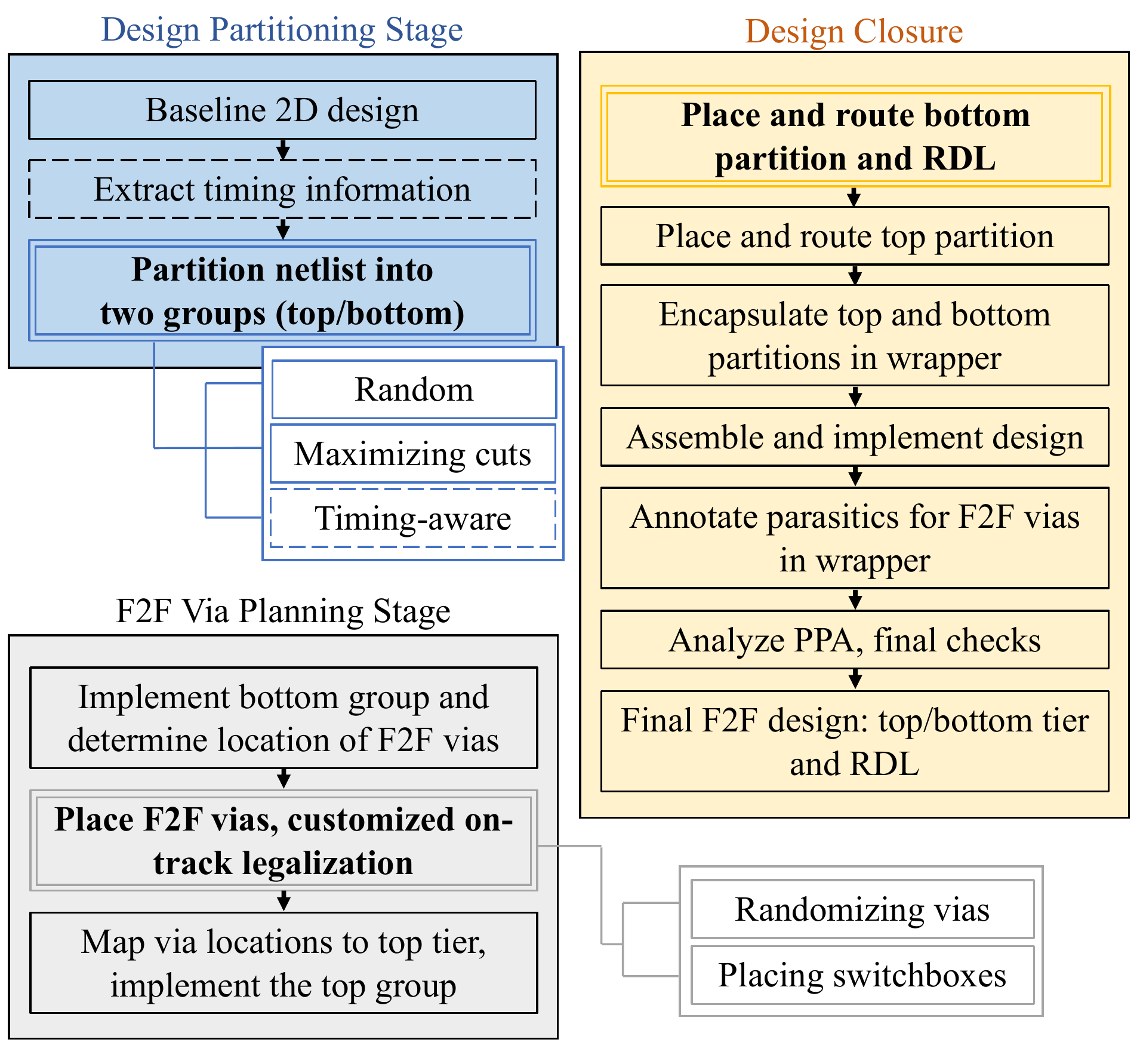}
\smallerspacecaption
\smallerspacecaption
\caption{Our CAD flow for F2F 3D ICs, implemented in \emph{Cadence Innovus}. Security-driven steps are emphasized in bold.
\label{fig:flow}
}
\smallerspacecaption
\smallerspacecaption
\smallerspacecaption
\end{figure}

Here we elaborate on the CAD and manufacturing flow for our notion of security-driven F2F stacking.
The CAD flow is in parts inspired by Chang \emph{et al.}~\cite{chang16}, but we devise our flow with a particular
focus on IP protection (Fig.~\ref{fig:flow}).
Our flow allows a concerned designer to explore the trade-offs between PPA and \emph{cuts}, i.e., the number of F2F inter-tier connections.
Cuts are a crucial metric for the security analysis, which is discussed in more detail in Sec~\ref{sec:security}.
It is also important to note that we follow the call for \emph{layout
anonymization}~\cite{imeson13}---we purposefully do not engage cross-tier optimization steps, to mitigate layout-level hints on the
obfuscated BEOL/RDLs.

As for the F2F process, we propose the following security-driven modification.
The wafers for the two tiers are fabricated by
one (two) untrusted foundry (foundries) and then shipped to a trusted BEOL and stacking facility. This trusted facility grows the obfuscated RDLs
on top of one wafer, and continues with the regular F2F flow (i.e., flipping and bonding the second wafer on top).

\subsection{Design Partitioning}
\label{sec:design_partitioning}

After obtaining the post-routed 2D design, we partition the netlist into \emph{top} and \emph{bottom} groups, representing the 
tiers of the F2F IC.
I/O ports are created for all interconnects between the two groups, representing the F2F vias.
Besides these F2F ports, we place primary I/Os at the chip boundary, as in conventional 2D designs. (This is also practical for F2F
		integration where TSVs are to be manufactured at the chip boundary for primary I/Os and the P/G grid.)

\textbf{Random partitioning:}
A naive way for security-driven partitioning is to assign gates to the top/bottom groups randomly. When doing so,
the number of cuts will be dictated by the number, type, and local interconnectivity of gates
being assigned to one group.

\textbf{Maximizing the cut-size:}
As already indicated (and further explored in Sec~\ref{sec:security}), the larger the cut size, the more difficult becomes IP piracy. Hence,
   here we seek to increase the cut size as much as reasonably possible.
First, timing reports for the 2D baseline are obtained. Next, gates are randomly alternated along their timing paths towards the top/bottom
groups.  In the security-wise best case (which is also the worst-case regarding power and performance), every other gate is assigned to
the top and bottom group, respectively; for a path with $n$ gates, $2n$ cuts are arising.

\textbf{Timing-aware partitioning:}
Here we seek to reduce layout cost while maintaining strong protection.
First, the available timing slack is determined for each gate.
Then, based on a user-defined threshold, the critical gates remain in the bottom tier, whereas all other gates are moved to the top tier.
This procedure is repeated with revised timing thresholds until an even utilization for both tiers is achieved.
Note that it is not easy for an attacker to understand whether any path in the
bottom/top group is critical or not (or complete, for that matter).
In other words, the attacker has to tackle both groups at once and, more importantly, resolve the randomized F2F vias and the obfuscated interconnects (see below).

\subsection{Planning of F2F Interconnects}

After placing the bottom tier, the initial locations for F2F ports are determined in the vicinity of the drivers/sinks.  Then, a
security-driven, i.e., randomized placement of F2F ports is conducted, along with customized on-track legalization.
Next, \emph{obfuscated switchboxes} are placed, and the
F2F ports are mapped to the top tier.

\begin{figure}[tb]
\smallerspacecaption
\centering
\includegraphics[width=.52\columnwidth]{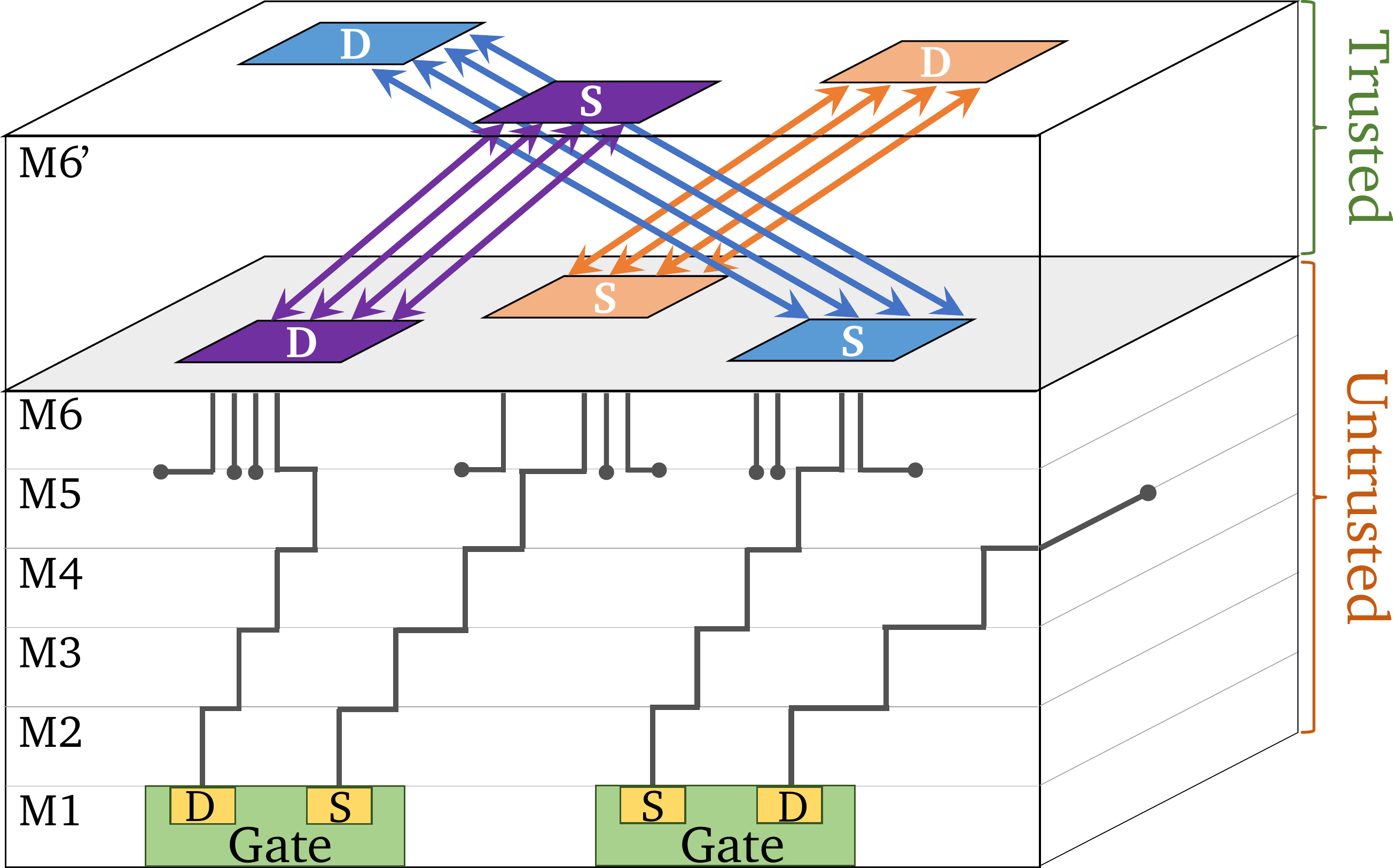}\hfill
\includegraphics[width=.45\columnwidth]{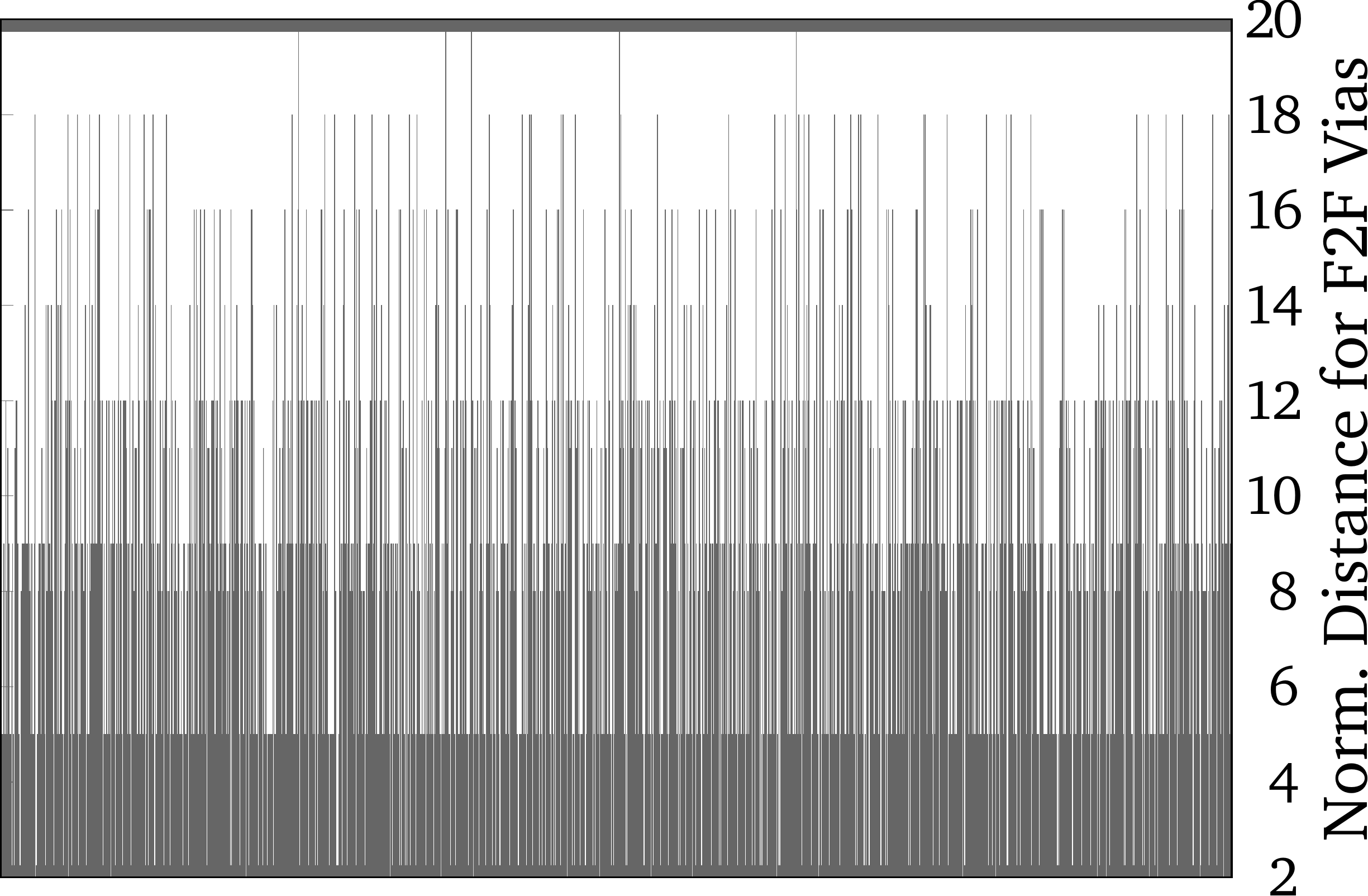}
\smallerspacecaption
\caption{(Left) RDL randomization for switchboxes and F2F vias.
	(Right) Normalized distances between to-be-connected F2F vias after randomization, for benchmark \emph{b17\_1}.
\label{fig:random_F2F}
}
\smallerspacecaption
\smallerspacecaption
\end{figure}

\textbf{Randomization:} It is easy to see that regular planning of F2F interconnects cannot be secure,
as this aligns the ports for the bottom and top tier directly. 
Hence, we randomize the arrangement of F2F ports as follows.
(Fig.~\ref{fig:random_F2F}).  We place additional F2F ports randomly (yet
		with the help of the on-track
		legalization) in the top RDL.
These randomized ports are then routed through the RDLs towards the original F2F ports
	connecting with the bottom tier, which are also embedded into custom switchboxes (see next).

\begin{figure}[tb]
\smallerspacecaption
\centering
\includegraphics[width=.92\columnwidth]{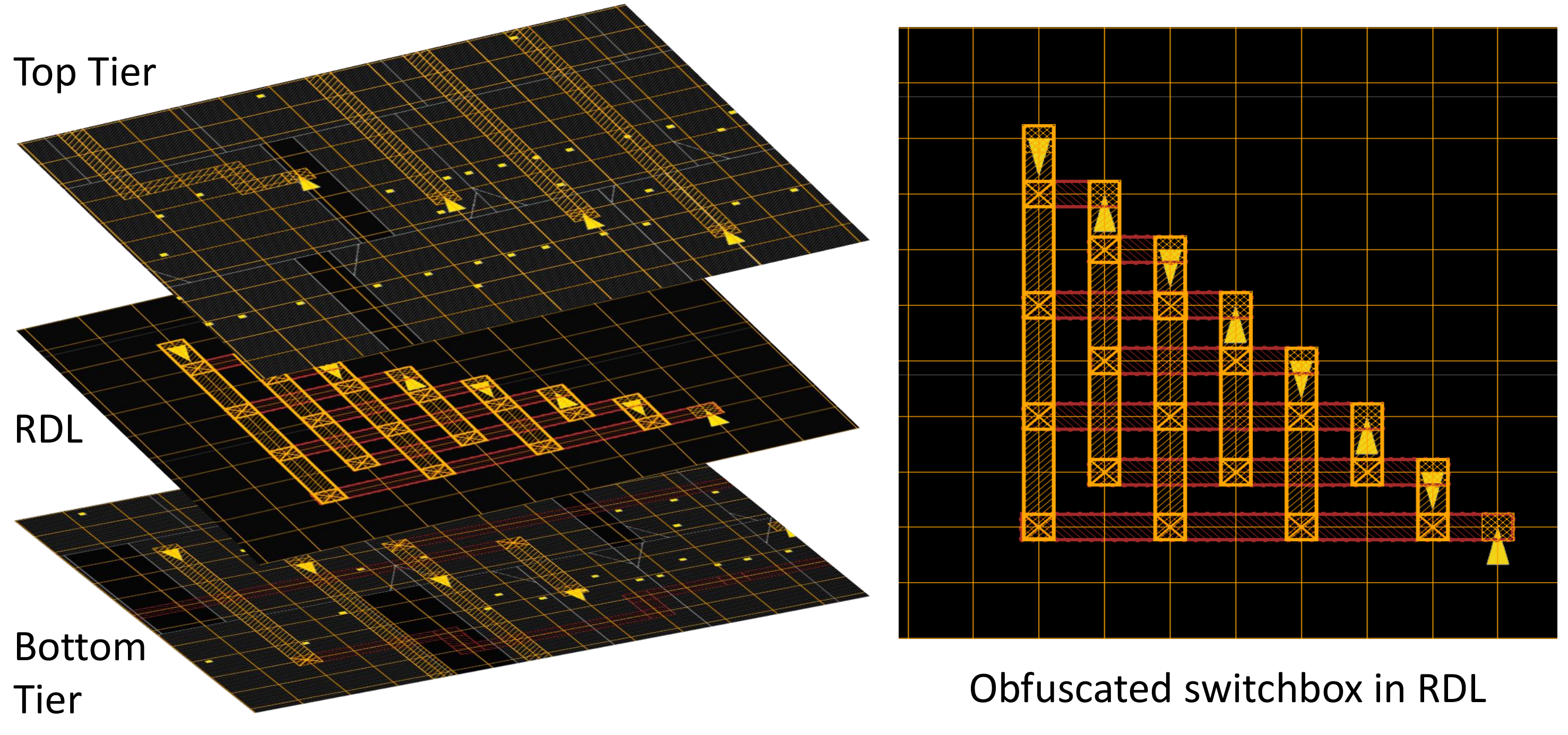}
\smallerspacecaption
\smallerspacecaption
\caption{Obfuscated switchbox, exemplarily for bottom-to-top drivers.
	Each driver pin (downwards triangle) can connect to any sink pin (upwards triangle).
	All F2F ports are aligned with the pins of the switchbox here, for simplicity,
	    whereas the top-tier ports are randomized in reality.
\label{fig:switchbox}
\smallerspacecaption
}
\end{figure}

\textbf{Obfuscated switchboxes:}
	To protect against reverse engineering, we obfuscate the
connectivity in the RDLs
using
a custom switchbox (Fig.~\ref{fig:switchbox}). This switchbox allows stealthy
one-to-one mapping of four drivers to four sinks.
The essence of the switchbox are Mg/MgO vias~\cite{chen15}, to cloak which driver
connects to which sink.
The pins of the switchbox represent the F2F ports.
To enable proper utilization of routing
resources, the pins are aligned with the routing tracks.
For randomization, the additional ports
connecting with the top tier are used for rerouting during design closure.

\textbf{On-track legalization:}
Each F2F port is moved inside the core boundary, towards the center point defined by all instances connected with this port.
Next, we obtain the closest and still-unoccupied on-track locations for actual placement.
If need be, we stepwise increase the search radius considering a user-defined threshold.

\subsection{Design Closure}
\label{sec:closure}

After the F2F via planning stage, both tiers are placed and routed separately.
Here we do not engage in any cross-tier optimization, to anonymize the individual tiers
from each other, but we apply intra-tier optimization.
While routing the bottom tier, we also route the randomized
and obfuscated RDL with their switchboxes. 
Next, we encapsulate 
     the top and bottom partitions in a \emph{wrapper netlist}, and we
assemble and implement the design followed by generating a SPEF
	file that captures the RC parasitics
of the F2F vias.
Finally, we perform DRC checks, evaluate the PPA, and stream out separate DEF files for the top/bottom tiers and the RDL.

\section{Results}
\label{sec:results}

\subsection{Experimental Setup}
\label{sec:setup}

\textbf{Implementation and layout evaluation:}
Our CAD flow is based on \emph{Cadence Innovus 17.1}, using custom \emph{Tcl} and \emph{Python} scripts, which impose negligible runtime
overheads.
We use the \emph{Nangate 45nm} library~\cite{nangate11} for our experiments, with six metal layers for the
baseline 2D setup and six layers for the top and bottom tier each in the F2F setup.
The RDL comprises four duplicated layers of M6, and F2F vias are modeled as M6 vias.
(While this is an optimistic assumption, for now,
F2F scaling can be expected to reach such dimensions.)
The PPA analysis is conducted for the slow process corner at 0.95V V$_{\textrm{DD}}$. 
For power analysis, we assume a switching activity of 0.2 for
all primary inputs.
We ensure that the layouts are free of any congestion, by choosing appropriate utilization rates. 
All experiments are carried out on an
Intel Xeon E5-4660 @ 2.2 GHz with \emph{CentOS 6.9}. For \emph{Cadence Innovus}, up to 16 cores are allocated.

\textbf{Setup for security evaluation:}
Since we promote 3D SM, regular proximity attacks such as~\cite{wang18_SM,magana17} cannot be applied.
Thus, we propose (and publicly release~\cite{webinterface}) a novel attack against 3D SM,
also accounting for the RDL obfuscation underlying in our scheme;
see also Sec.~\ref{sec:security}.
The strength of our attack is evaluated by commonly used metrics, i.e., the \emph{correct connection rate (CCR)} and \emph{Hamming distance (HD)}.
HD is calculated using \emph{Synopsys VCS} with 1,000,000 test patterns.
As for
SAT-based reverse engineering attacks, we leverage~\cite{subramanyan15}.
The related time-out is set to 72 hours.

\textbf{Designs:}
Benchmarks from the \emph{ISCAS-85} and \emph{ITC-99} suites are used for layout and security analysis.

\subsection{Security-Driven Layout Evaluation}
\label{sec:layout_PPA}

\begin{figure}[tb]
\centering
\includegraphics[width=.99\columnwidth]{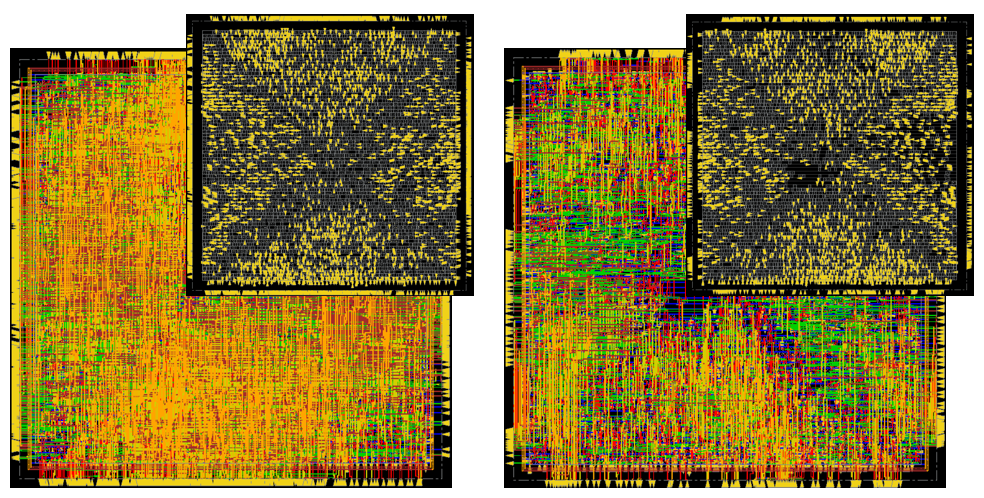}
\smallerspacecaption
\smallerspacecaption
\caption{Layout snapshots of bottom/top tier (left/right) for \emph{b22}. The insets show the corresponding F2F vias.
\label{fig:die_images}
}
\smallerspacecaption
\smallerspacecaption
\end{figure}

Our flow allows to trade off PPA and cuts; the latter dictates the resilience against IP piracy both during and
after manufacturing.
Figure~\ref{fig:die_images} showcases the layout images for
benchmark \emph{b22}.

\begin{figure*}[tb]
\centering
	\captionsetup[subfigure]{labelformat=empty}
\smallerspacecaption
\smallerspacecaption
\smallerspacecaption
\smallerspacecaption
\subfloat[]{\includegraphics[width=.99\textwidth]{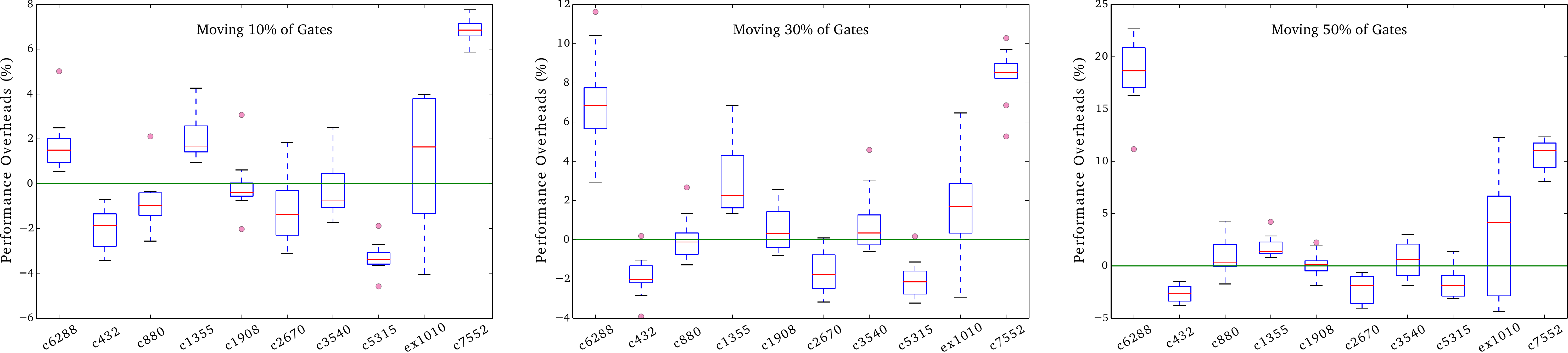}}\\
\smallerspacecaption
\smallerspacecaption
\smallerspacecaption
\smallerspacecaption
\subfloat[]{\includegraphics[width=.99\textwidth]{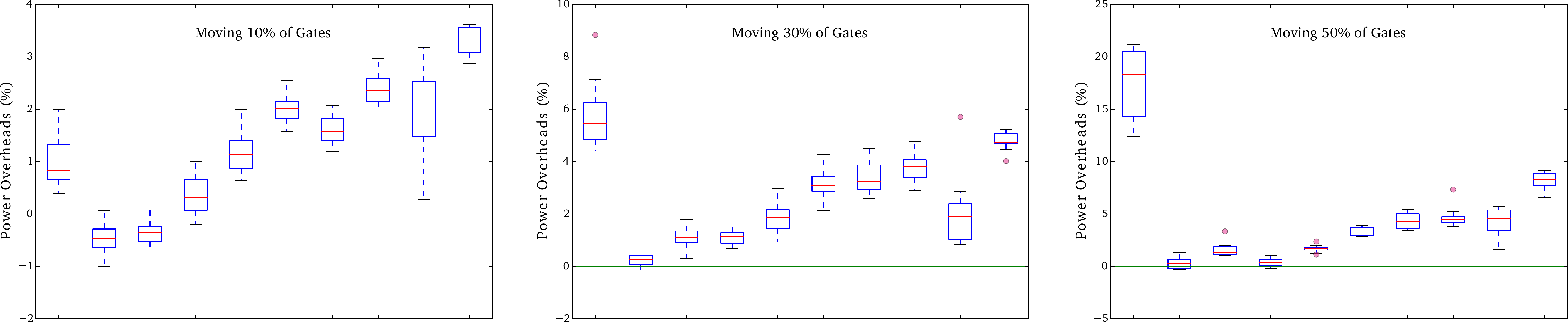}}
\smallerspacecaption
\smallerspacecaption
\smallerspacecaption
\smallerspacecaption
\caption{Impact of maximizing the cuts or F2F vias, by moving of gates, on performance (top) and power (bottom). Each boxplot represents ten
	runs.
\label{fig:PPA_MC}
\smallerspacecaption
}
\end{figure*}

\textbf{Maximizing the cut-size:}
Here we move gates
from the bottom to the top group
in steps of 10\%, up to 50\%.
As the strategy is randomized, we perform ten runs for each benchmark for any given percentage of gates to move.
The resulting power and performance distributions are illustrated
in Fig.~\ref{fig:PPA_MC}.
Interestingly, even for the security-wise best case of randomly moving 50\% of the gates,
some runs still provide better power and/or performance than the 2D baseline.
This finding demonstrates the potential for our scheme.
Note that we refrain both from randomizing the F2F ports
and from using the obfuscated switchboxes for these initial experiments.

\begin{figure}[tb]
\centering
\includegraphics[width=.99\columnwidth]{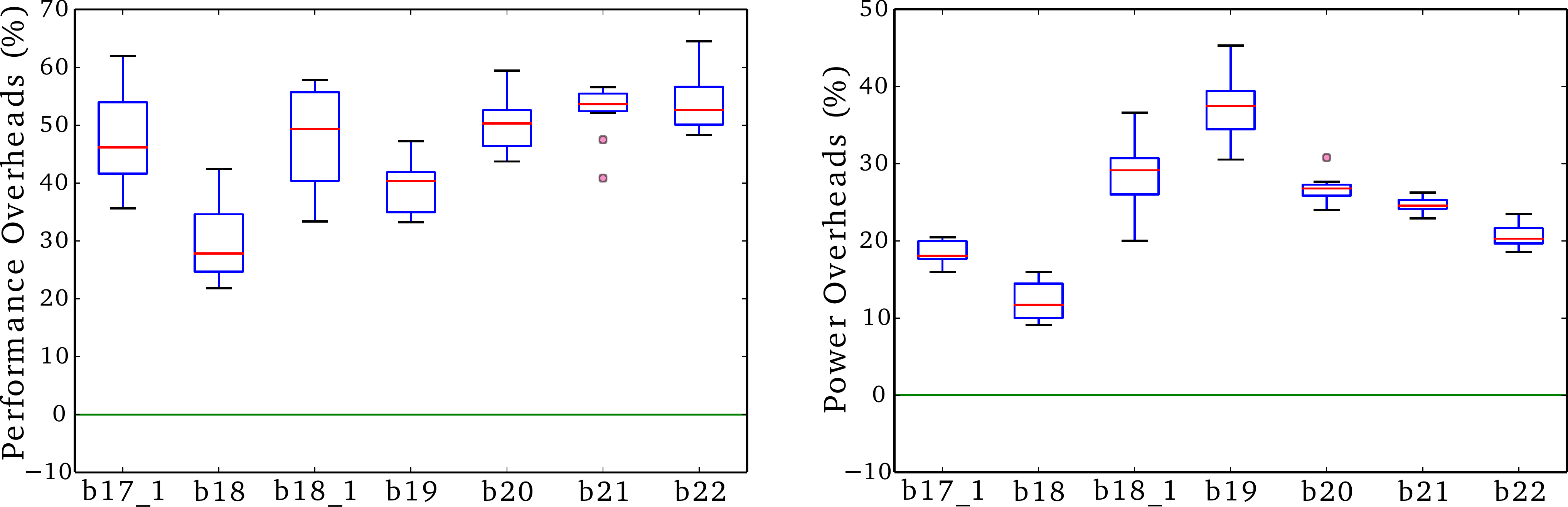}
\smallerspacecaption
\caption{Layout cost for maximizing cuts, 35--50\% gates moved, obfuscated switchboxes and F2F randomization. Each boxplot represents ten
	runs.
\label{fig:MC_plus_pin_random_cross}
}
\smallerspacecaption
\smallerspacecaption
\smallerspacecaption
\end{figure}

Once F2F ports are randomized and switchboxes are used, larger benchmarks such as \emph{b18\_1} may incur overheads of up to
60\% (Fig.~\ref{fig:MC_plus_pin_random_cross}).  Hence, although this strategy offers strong resilience,
a more aggressive PPA-security trade-off may be desired.

\begin{figure}[tb]
\centering
\includegraphics[width=.99\columnwidth]{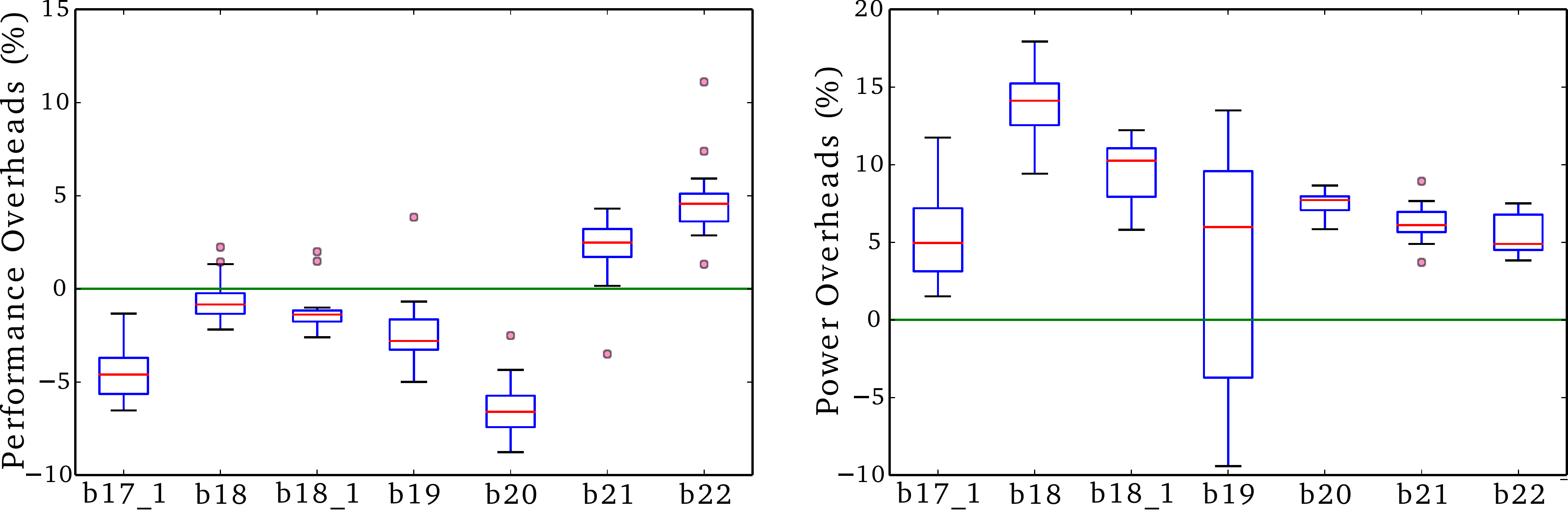}
\smallerspacecaption
\caption{Performance, power cost for timing-aware 3D setup with obfuscated switchboxes and F2F randomization. Each boxplot represents ten
	runs.
\label{fig:TA_plus_pin_random_cross}
}
\smallerspacecaption
\end{figure}

\textbf{Timing-aware partitioning with F2F randomization and obfuscated switchboxes:}
This setup tackles the need outlined above.
In fact, we observe that even for larger benchmarks (Fig.~\ref{fig:TA_plus_pin_random_cross}), there can be
some
layout benefits when comparing these 3D designs to their 2D baseline.
To demonstrate the
security implication of this setup, we plot the normalized distances between to-be-connected F2F vias in
Fig.~\ref{fig:random_F2F}. This figure shows a wide variation across the inter-tier nets, whereas for regular/unprotected F2F stacking the
distances would be all zero.
For a more detailed security analysis, see Sec~\ref{sec:security}. Next,
we compare our work to prior art.

\textbf{Comparison with LC schemes:}
Threshold-based LC is recently gaining traction. Although promising in terms of resilience (for some schemes even during manufacturing), the
PPA overheads are considerable.  For example, Akkaya \emph{et al.}~\cite{Akkaya2018} report overheads of 9.2$\times$,
    6.6$\times$, and 3.3$\times$ for PPA, respectively, when compared to conventional 2-input NAND gate.
	    Nirmala~\emph{et al.}~\cite{nirmala16} report 11.2$\times$ and 10.5$\times$ cost for power and area, respectively.
    Collantes~\emph{et
	    al.}~\cite{collantes16} report power and performance cost of 72\% and 31\%, respectively, for 40\% camouflaging.
In~\cite{DBLP:journals/corr/JangG17}, threshold voltages are leveraged to obfuscate the interconnects, leading to PPA overheads of 29\%,
	44\%, and 33\%, namely when 15\% of the nets are obfuscated.
In~\cite{patnaik17_Camo_BEOL_ICCAD}, PPA overheads of 4.9\%, 31.2\%, and 25\%
are reported for \emph{b17} at 60\% LC (by obfuscation of the interconnects).
Even when compared to the latter more promising schemes, we can provide
significantly better PPA (except for~\cite{patnaik17_Camo_BEOL_ICCAD} concerning power).

Dofe, Yan, \emph{et al.}~\cite{dofe16_mono3D,yan17_camo} recently proposed LC for monolithic 3D ICs. At the time of writing, their libraries
were not available to us for a detailed comparison. More importantly, however, manufacturing of such camouflaged 3D ICs requires trust into an
advanced fab. The notion of 3D SM as in our scheme cannot be applied for monolithic 3D ICs (due to the sequential manufacturing process)
	and, thus, their scheme~\cite{dofe16_mono3D,yan17_camo} \emph{cannot} protect the IP at manufacturing
time.

\textbf{Comparison with SM schemes:}
In Table~\ref{tab:sec-comp2}, we compare with some studies on 2D SM.
Overall, the placement-centric techniques by Wang \emph{et al.}~\cite{wang18_SM}
are competitive 
concerning
power and performance. However, as is always the case for regular SM,
Wang \emph{et al.} can only avert fab-based adversaries,
\emph{but not malicious end-users}.

\begin{table*}[tb]
\centering
\footnotesize
\caption{ 
PPA cost comparison with 2D SM protection schemes. Numbers are in \% and quoted from the respective publications.
}
\smallerspacecaption
\smallerspacecaption
\begin{tabular}{|*{16}{c|}}
\hline
\multirow{2}{*}{\textbf{Benchmark}}
& \multicolumn{3}{|c|}{\textbf{BEOL+Physical~\cite{wang18_SM}}} 
& \multicolumn{3}{|c|}{\textbf{Logic+Physical~\cite{wang18_SM}}} 
& \multicolumn{3}{|c|}{\textbf{Logic+Logic~\cite{wang18_SM}}} 
& \multicolumn{3}{|c|}{\textbf{Concerted Lifting~\cite{patnaik_ASPDAC18}}} 
& \multicolumn{3}{|c|}{\textbf{Proposed with Random Partitioning}} \\
\cline{2-16}
& \textbf{Area} & \textbf{Power} & \textbf{Delay}
& \textbf{Area} & \textbf{Power} & \textbf{Delay}
& \textbf{Area} & \textbf{Power} & \textbf{Delay}
& \textbf{Area} & \textbf{Power} & \textbf{Delay}
& \textbf{Area$^*$} & \textbf{Power} & \textbf{Delay}
\\ \hline
c432 & N/A & 0.17 & 0.49 &
N/A & 0.44 & 0.24 &
N/A & 0.17 & 0.21 &
7.7 & 13.1 & 11.6 &
-50 & -2.66 & 0.31
 \\ \hline
 
c880 & N/A & 0.25 & 0.05 &
N/A & 0.35 & 0.03 &
N/A & -0.05 & -0.09 &
0 & 12.1 & 19.9 &
-50 & 0.97 & 1.6
 \\ \hline
 
c1355 & N/A & 0.52 & 0.57 &
N/A & 0.75 & 0.42 &
N/A & 0.03 & 0.01 &
0 & 12.2 & 21.3 &
-50 & 1.83 & 0.38
\\ \hline

c1908 & N/A & 1.1 & 1.3 &
N/A & 1.1 & 0.23 &
N/A & 0.45 & 0.39 &
7.7 & 14.6 & 18.9 &
-50 & 0.11 & 1.69
\\ \hline

c2670 & N/A & 0.29 & 0.27 &
N/A & 0.29 & 0.27 &
N/A & 0.05 & 0.03 &
7.7 & 10 & 12 &
-50 & -2.18 & 3.32
\\ \hline

c3540 & N/A & 0.53 & 0.28 &
N/A & 0.36 & 0.02 &
N/A & 0.14 & -0.02 &
7.7 & 5 & 2.8 &
-50 & 0.59 & 4.32
\\ \hline

c5315 & N/A & 0.19 & -0.01 &
N/A & 0.67 & 0.08 &
N/A & 0.29 & -0.01 &
7.7 & 7.9 & 16.9 &
-50 & -1.66 & 4.73
\\ \hline

c6288 & N/A & 0.29 & 0.19 &
N/A & 0 & 0 &
N/A & 0.1 & 0.67 &
27.3 & 12.3 & 15.7 &
-50 & 10.43 & 10.21
\\ \hline

c7552 & N/A & 0.28 & -0.36 &
N/A & 0.35 & -0.05 &
N/A & 0.56 & 1.77 &
16.7 & 9.3 & 15.7 &
-50 & 10.57 & 8.21
\\ \hline

\textbf{Average} & N/A & 0.4 & 0.31 &
N/A & 0.48 & 0.14 &
N/A & 0.19 & 0.33 & 
9.2 & 10.7 & 15 &
-50 & 2 & 3.86
\\ \hline

\end{tabular}
\\ \footnotesize
$^*$Following the standard practice for 3D studies, we report on area by considering individual die outlines.
In~\cite{patnaik_ASPDAC18}, area is reported in terms of die outlines as well.

\label{tab:sec-comp2}
\smallerspacecaption
\end{table*}

In Table~\ref{tab:comparison_with_2.5D}, we compare with the security-driven 2.5D integration scheme by Xie \emph{et al.}~\cite{xie17}.
Their work is relevant as they
propose a similar notion of security based on cut sizes. For the benchmarks the authors
considered, we obtain on average 53\% more cuts in our scheme. (For our cut sizes for larger benchmarks, refer to
		Table~\ref{tab:security_analysis}).
Regarding
PPA, we observe significantly lower costs than~\cite{xie17}.\footnote{\label{fn:die_outlines}
	Concerning area, note that we report on die outlines, which is
	standard practice for 3D studies.
Accordingly, for our numbers of -50\%, the F2F 3D IC and the 2D baseline require the same total silicon area, i.e., we incur 0\%
absolute area cost.
While Xie \emph{et al.} report similar cost,
they omit that their scheme requires
an interposer which---being at least as large as the chips stacked onto it---incurs $\geq$100\% cost.
Still, mainly comprising metal layers, we acknowledge that an interposer is less expensive than regular chips.}
Besides, as with regular SM, their 2.5D scheme is 
\emph{not} inherently
   resilient against malicious end-users, but our 3D scheme is.

\subsection{Security Analysis and Attacks}
\label{sec:security}

\textbf{Proximity attack for 3D SM:}
To the best of our knowledge, there is no attack yet in the literature which can account for 3D SM.
Hence, we propose and implement such an attack, with a focus on one untrusted foundry (or two colluding foundries) and our RDL obfuscation. We
provide this attack as a public release in~\cite{webinterface}.

We assume that the attacker holds the layout files for the top and bottom tier,
but \emph{initially} she has no access to the trusted RDL
		(we discuss the implications for obtaining the RDL further below).
Although she understands how many drivers are connecting from the bottom to the top tier and vice versa, 
she does not know which driver connects to which sink, given the randomized F2F vias.
Recall that we do not engage in cross-tier optimization, to mitigate any layout-level hints.\footnote{Also recall
the different-foundries scenario in Sec.~\ref{sec:diff_found}, which is significantly more challenging. There, the attacker has not only to
tackle the driver-sink mappings but furthermore guess the set of gates withhold by the trusted foundry.}
Let us assume there are $d_{bot}$ drivers in the bottom and $d_{top}$ drivers in the top tier.
Since we do not allow for fan-outs within the RDL (as this would occupy more F2F vias than necessary), there are only one-to-one
mappings---this results in $d_{bot}! \times d_{top}!$ possible netlists.
Once switchboxes are used, however,
the attacker can tackle groups of four drivers/sinks at once.
Still, she has to resolve (a)~which four top-tier drivers are connected to which four bottom-tier sinks and vice versa, and (b) the
connectivity within the obfuscated switchboxes.
For those cases, there are
$4!\times \left(\left(1/4 \times d_{bot}\right)! \times \left(1/4 \times d_{top}\right)!\right)$ possible netlists remaining.
Next, we outline the corresponding heuristics at the heart of our attack.

\begin{enumerate}
\item \emph{Unique mappings:} Any driver in the bottom/top tier will feed only one sink in the top/bottom tier.
Hence, an attacker will reconnect drivers and sinks individually.
Moreover, she can identify all primary I/Os as they are
implemented using wirebonds or TSVs, not randomized F2F vias.
\item \emph{Layout hints:} Although the F2F vias are randomized, the attacker may try to correlate the proximity and
orientation of F2F vias with their corresponding but withheld RDL connectivity.
Towards this end, she can also leverage the routing towards 
the switchbox ports.
Moreover, recalling the practical threat model, the attacker may be able to identify some known IP and accordingly confine the related sets
of candidate F2F interconnects.
Our attack is generic and can account for those scenarios, by keeping track of the candidate F2F pairings considered by the attacker.
\item \emph{Combinatorial loops:} Since both tiers and thus all active components are available to the attacker, she can readily exclude
those F2F connections inducing combinatorial loops.
\end{enumerate}

We provide empirical attack results in Table~\ref{tab:security_analysis}.
Here we assume that the attacker was able to correctly infer all the driver-sink pairings through 
the switchboxes, only the obfuscation within switchboxes themselves remains to be attacked.
This is a \emph{strong} assumption and rendering our evaluation conservative.
In fact, this scenario can be considered as an optimal proximity attack, as
for all F2F connections the correct one is always among the considered candidates.
The results in Table~\ref{tab:security_analysis} indicate the computational efficiency of our attack for smaller designs, but also the challenges once larger designs with 
large solution spaces are to be tackled.
With regards to CCR and HD for the successfully recovered netlists, our protection scheme can be considered as reasonably secure.

\begin{table}[tb]
\centering
\scriptsize
\setlength{\tabcolsep}{0.55em}
\caption{Comparison with~\cite{xie17}. PPA is in contrast to a 2D baseline, numbers are in \%.
	See also Footnote~\ref{fn:die_outlines} on area.
}
\smallerspacecaption
\smallerspacecaption
\begin{tabular}{|c|c|c|c|c|c|c|c|c|}
\hline
\multirow{2}{*}{\textbf{Benchmark}} 
& \multicolumn{4}{|c|}{\textbf{Xie \emph{et al.}~\cite{xie17} (SC+SP)}} 
& \multicolumn{4}{|c|}{\textbf{Proposed with Random Partitioning}} \\
\cline{2-9}
 &
 \textbf{Cut Size} &
 \textbf{Area} &
 \textbf{Power} &
 \textbf{Delay} &
 \textbf{Cut Size} &
 \textbf{Area} &
 \textbf{Power} &
 \textbf{Delay} \\
 \hline \hline
c432 &  
130 & 1  & 17.6  & 5.9  & 134  & -50 (0) & -2.66 & 0.31 \\ \hline
c880 &  
141 & 0  & 29.4  & 10  & 138  & -50 (0) & 0.97 & 1.6 \\ \hline
c1355 &  
130 & 0  & 17.6  & 17.6 & 91  & -50 (0) & 1.83 & 0.38 \\ 
\hline
c1908 &  
132 & 1  & 11.8  & 29.4 & 149  & -50 (0) & 0.11 & 1.69 \\
\hline
c2670 &  
152 & 0  & 11.8  & 5.9 & 154  & -50 (0) & -2.18 & 3.32 \\
\hline
c3540 &  
133 & 0  & 5.9  & 5.9 & 349  & -50 (0) & 0.59 & 4.32 \\
\hline
c7552 &  
157 & 1  & 1  & 5.9 & 477  & -50 (0) & 10.57 & 8.21 \\
\hline
\textbf{Average} &  
139 & 0.4  & 13.6  & 11.5 & 213  & -50 (0) & 1.32 & 2.83 \\
\hline
\end{tabular}
\label{tab:comparison_with_2.5D}
\smallerspacecaption
\end{table}

\begin{table}[tb]
\centering
\scriptsize
\setlength{\tabcolsep}{0.17em}
\caption{Attack results on average. Time-out `t-o' is 72 hours.
}
\smallerspacecaption
\smallerspacecaption
\begin{tabular}{|c|c|c|c|c|c|c|}
\hline
\multirow{2}{*}{\textbf{Benchmark}} 
& \multicolumn{2}{|c|}{\textbf{Cut Sizes}}
& \textbf{SAT Attack~\cite{subramanyan15}}
& \multicolumn{3}{|c|}{\textbf{Proposed 3D-SM Proximity Attack}} \\
\cline{2-7}
 &
 \textbf{Random} &
 \textbf{Timing-Aware} &
 \textbf{Runtime (Min.)} &
 \textbf{CCR (\%)} &
 \textbf{HD (\%)} &
 \textbf{Runtime (Sec.)} \\
 \hline \hline
c432 & 134 & 56 &  624 & 30.4  & 45.2 & 10 \\ \hline
c880 & 138 & 53 &  642 & 27.8  & 39.4 & 23 \\ \hline
c1355 & 91 & 37 &  492 & 31.1  & 43.8 & 53 \\ \hline
c3540 & 349 & 97 &  948 & 22.6  & 41.3 & 3,729 (62 Minutes) \\ \hline
\emph{b17\_1} & 6,650 & 2,482 &  t-o & N/A  & N/A & t-o  \\ \hline
\emph{b18} & 15,974 & 6,906 &  t-o & N/A & N/A & t-o \\ \hline
\emph{b18\_1} & 16,706 & 6,616 &  t-o & N/A  & N/A & t-o \\ \hline
\emph{b19} & 33,417 & 13,142 &  t-o & N/A  & N/A & t-o \\ \hline
\end{tabular}
\label{tab:security_analysis}
\smallerspacecaption
\end{table}

\textbf{SAT-based attack:}
After manufacturing, the attacker can readily understand which four drivers/sinks are connected through the switchboxes, but she still has
to resolve the obfuscation within the switchboxes themselves.
The attacker may now leverage a working copy as an oracle and launch a SAT attack.
Towards that end, we employ the attack proposed in~\cite{subramanyan15}, and we model the problem using multiplexers as outlined in~\cite{massad15,yu17}.
Empirical results are given in Table~\ref{tab:security_analysis}. As expected, the SAT attack succeeds for smaller designs but runs
into time-out for larger designs. This finding is also consistent with those reported by Xie \emph{et al.}~\cite{xie17} for their security-driven
2.5D scheme, which has a security notion similar to our work.

\section{Conclusion and Outlook}
\label{sec:conclusion}

Initially, we review prior art
and their limitations.
We also put forward a novel, practical threat model of IP piracy which is in line with the business models of present-day chip companies.
Next, we elaborate in detail how 3D integration is a naturally strong match to
	combine SM and LC. (This also allows us to extend the defense scope of SM to practical commercial applications.)
Towards this end, we propose a security-driven CAD and manufacturing flow for face-to-face (F2F) ICs, an up-and-coming option for 3D
integration.
We conduct comprehensive experiments on DRC-clean layouts, and strengthened by an extensive security
analysis, we argue that entering the third dimension is promising for 
IP protection.

As for future work,
      we aim for a more formal method for partitioning gates across tiers, 
also to protect against other threats such as hardware Trojans.
In the broader sense,
we plan to explore if and how 3D integration can provide resilience against physical attacks such as invasive probing or
exploitation of side-channel leakage.
 
\section*{Acknowledgments}
\label{sec:acknowledgments}

This work was supported in part by the Center for Cyber Security (CCS) at NYU New York/Abu Dhabi (NYU/NYUAD).
We also thank Dr.\ Anja Henning-Knechtel for preparing selected illustrations.

\bibliographystyle{IEEEtran}
\balance\bibliography{main} 

\end{document}